\documentclass[final,3p,times,epsfig,colordvi,twocolumn]{elsart5p}
\usepackage{amsmath}

\usepackage{graphics}
\journal{Physics Letters B}

\begin{document}
\begin{frontmatter}
\title{Double pion photoproduction off nuclei - are there effects beyond final-state
interaction?
}
\author[Basel]{Y.~Maghrbi},
\author[Giessen]{R.~Gregor},  
\author[Giessen]{S.~Lugert},   
\author[Mainz]{J.~Ahrens},             
\author[Glasgow]{J.R.M.~Annand},          
\author[Mainz]{H.J.~Arends},
\author[Bonn]{R.~Beck},
\author[Petersburg]{V.~Bekrenev},         
\author[Basel]{B.~Boillat}, 
\author[Pavia]{A.~Braghieri},
\author[Edinburgh]{D.~Branford},            
\author[Washington]{W.J.~Briscoe},
\author[UCLA]{J.~Brudvik},  
\author[Lebedev]{S.~Cherepnya},
\author[Glasgow]{R.Codling},              
\author[Mainz,Glasgow,Washington]{E.J.~Downie},            
\author[Giessen]{P.~Dexler},
\author[Lebedev]{L.V.~Fil'kov},
\author[Tomsk]{A.~Fix},              
\author[Edinburgh]{D.I.~Glazier},
\author[Mainz]{E.~Heid},                
\author[Sackville]{D.~Hornidge},            
\author[Mainz]{O.~Jahn},
\author[Basel]{I.~Jaegle},
\author[Lebedev,Mainz]{V.L.~Kashevarov},
\author[Basel]{I. Keshelashvili}, 
\author[Zagreb]{A.~Knezevic},
\author[Moskau]{A.~Kondratiev},
\author[Zagreb]{M.~Korolija},          
\author[Basel,Giessen]{M.~Kotulla},
\author[Mainz]{D.~Krambrich},
\author[Basel]{B.~Krusche}\ead{Bernd.Krusche@unibas.ch}, 
\author[Mainz]{M. Lang},
\author[Moskau]{V.~Lisin},         
\author[Glasgow]{K.~Livingston},    
\author[Glasgow]{I.J.D.~MacGregor},
\author[Kent]{D.M.~Manley},
\author[Mainz]{M.~Martinez},                  
\author[Glasgow]{J.C.~McGeorge},
\author[Zagreb]{D.~Mekterovic},          
\author[Giessen]{V.~Metag},
\author[UCLA]{B.M.K.~Nefkens},         
\author[Bonn]{A.~Nikolaev},   
\author[Giessen]{R.~Novotny},
\author[Mainz]{M.~Ostrick},
\author[Pavia]{P.~Pedroni},
\author[Basel]{F.~Pheron},
\author[Moskau]{A.~Polonski},
\author[UCLA]{S.N.~Prakhov},
\author[UCLA]{J.W.~Price}, 
\author[Glasgow]{G.~Rosner},
\author[Mainz]{M.~Rost},
\author[Basel,Pavia]{T.~Rostomyan},
\author[Giessen]{S.~Schadmand},            
\author[Mainz]{S.~Schumann},
\author[Catholic]{D.~Sober},
\author[UCLA]{A.~Starostin},  
\author[Zagreb]{I.~Supek},  
\author[Mainz]{A.~Thomas},             
\author[Mainz]{M.~Unverzagt},                       
\author[Edinburgh]{D.P.~Watts},
\author[Basel]{D.~Werthm\"uller},
\author[Basel]{F.~Zehr}
\address[Basel] {Department of Physics, University of Basel, Ch-4056 Basel, Switzerland}
\address[Giessen] {II. Physikalisches Institut, University of Giessen, D-35392 Giessen, Germany}
\address[Mainz] {Institut f\"ur Kernphysik, University of Mainz, D-55099 Mainz, Germany}
\address[Glasgow] {SUPA, School of Physics and Astronomy, University of Glasgow, G12 8QQ, United Kingdom}
\address[Bonn] {Helmholtz-Institut f\"ur Strahlen- und Kernphysik, University Bonn, D-53115 Bonn, Germany}  
\address[Petersburg] {Petersburg Nuclear Physics Institute, RU-188300 Gatchina, Russia}
\address[Pavia] {INFN Sezione di Pavia, I-27100 Pavia, Pavia, Italy}
\address[Edinburgh] {School of Physics, University of Edinburgh, Edinburgh EH9 3JZ, United Kingdom}
\address[Washington] {Center for Nuclear Studies, The George Washington University, Washington, DC 20052, USA}
\address[UCLA] {University of California Los Angeles, Los Angeles, California 90095-1547, USA}
\address[Lebedev] {Lebedev Physical Institute, RU-119991 Moscow, Russia}
\address[Tomsk] {Laboratory of Mathematical Physics, Tomsk Polytechnic University, Tomsk, Russia}
\address[Sackville] {Mount Allison University, Sackville, New Brunswick E4L1E6, Canada}
\address[Zagreb] {Rudjer Boskovic Institute, HR-10000 Zagreb, Croatia}
\address[Moskau] {Institute for Nuclear Research, RU-125047 Moscow, Russia}
\address[Kent] {Kent State University, Kent, Ohio 44242, USA}
\address[Catholic] {The Catholic University of America, Washington, DC 20064, USA}

\begin{abstract}
Photoproduction of $\pi^{0}\pi^{0}$ and $\pi^{0}\pi^{\pm}$ pairs from nuclei
has been measured over a wide mass range ($^2$H, $^{7}$Li, $^{12}$C, $^{40}$Ca,
and $^{\rm nat}$Pb) for photon energies from threshold to 600 MeV. The 
experiments were performed at the MAMI accelerator in Mainz, using the Glasgow photon 
tagging spectrometer and a 4$\pi$ electromagnetic calorimeter consisting of the 
Crystal Ball and TAPS detectors. A shift of the pion-pion invariant mass spectra for heavy 
nuclei to small invariant masses has been observed for $\pi^0$ pairs but also for the
mixed-charge pairs. The precise results allow for the first time a model-independent 
analysis of the influence of pion final-state interactions. The corresponding effects 
are found to be large and must be carefully considered in the search for possible
in-medium modifications of the $\sigma$-meson.
Results from a transport model calculation reproduce the shape of the invariant-mass 
distributions for the mixed-charge pairs better than for the neutral pairs, but also 
for the latter differences between model results and experiment are not large, leaving
not much room for $\sigma$-in-medium modification.
\end{abstract}
\end{frontmatter}

\section{Introduction}

The generation of the mass of hadrons composed of light quarks is a central 
problem in quantum chromodynamics (QCD), the theory of the strong interaction. 
Unlike any other composite system, hadrons are built out of constituents with masses 
which are negligible compared to their total mass which is generated by dynamical 
effects. A central role is played by the spontaneous breaking of chiral symmetry, a 
fundamental symmetry of QCD. Without this symmetry breaking, hadrons would appear as 
mass degenerate parity doublets. However, in the spectrum of free particles large 
mass splitting is observed between chiral partners, for baryons and for mesons. 
The order parameter of the 
symmetry breaking, the quark condensate, is expected to be density and temperature 
dependent so that the symmetry could be at least partially restored for hadrons embedded 
in nuclear matter. Already in 1991 Brown and Rho \cite{Brown_91} suggested scaling laws 
for hadron in-medium masses; and shortly after, Lutz, Klimt, and Weise \cite{Lutz_92} 
discussed in-medium properties of mesons in the framework of the Nambu-Jona-Lasinio 
(NJL) model. A recent overview of theory and experiment is given in \cite{Leupold_10}; 
results from photon-induced reactions are summarized in \cite{Krusche_11}. Although 
originally in-medium effects were mainly discussed in connection with heavy ion collisions, 
they should be also observable at normal nuclear density \cite{Lutz_92} in systems 
that can be more easily interpreted than the strongly time-dependent density and 
temperature variations in heavy ion collisions. 

Among such effects is the mass-shift of the $\sigma$-meson in normal dense 
nuclear matter. This scalar $J^{P}=0^+$, isoscalar $I=0$ state is mostly regarded as 
the chiral partner of the $J^{P}=0^-, I=1$ pion \cite{Bernard_87,Hatsuda_99,Aouissat_99}. 
However, due to its unconventionally large width (mass between 400 - 550 MeV, width 
400 - 700 MeV \cite{PDG}), in some models it is treated as a correlated state of two pions 
(respectively four quarks) rather than a quark-antiquark state 
\cite{Chiang_98,Roca_02,Chanfray_06,Albaladejo_12}. The mass split between the $\sigma$ 
and the pion in vacuum is large. However, predictions from different models 
\cite{Bernard_87,Roca_02} indicated that already at normal nuclear matter density $\rho_0$
the mass of the $\sigma$ should drop by $\approx$ 200 MeV \cite{Bernard_87,Roca_02},
while the pion, which is protected by its Goldstone boson nature, should almost be unaffected. 
The strong coupling of the $\sigma$ to scalar, isoscalar pion pairs should lead 
to a modification of the invariant-mass distribution of such pairs in nuclear matter. 
This effect is predicted by different model studies independent of the assumed nature of 
the $\sigma$ \cite{Hatsuda_99,Aouissat_99,Chiang_98,Roca_02,Chanfray_06,Rapp_99}.
In models, in which the $\sigma$ is not treated as a quark-antiquark state, the 
corresponding effect comes from an in-medium modification of pion-pion final state 
interaction for scalar, isoscalar pion pairs. This effect must not be confused with
the final state interaction of individual pions with nuclear matter by
re-absorption and re-emission processes.   

This prediction has been experimentally tested with pion- and photon-induced
reactions. Pion induced reactions were studied by the CHAOS collaboration
\cite{Bonutti_96,Bonutti_99,Bonutti_00,Camerini_04,Grion_05}, which observed a mass
shift for isoscalar $\pi^+\pi^-$ pairs in heavy nuclei but no effect for the like-sign 
$\pi^+\pi^+$ pairs where the $\sigma$ cannot contribute. The experiment had only a limited 
detector acceptance, which complicated the interpretation of the results. A similar effect was observed for the $\pi^- A\rightarrow A\pi^0\pi^0$ reaction by the Crystal Ball Collaboration 
at BNL \cite{Starostin_00} which, however, could not measure an isovector channel for comparison.
Pion-induced reactions have the disadvantage that due to the large absorption
cross section only the nuclear surface is probed.

Photons can probe the entire volume of nuclei. Although also in this case final-state
interaction (FSI) effects cannot be avoided, they can be studied in a systematic way 
because they depend strongly on the kinetic energy of the produced pions \cite{Krusche_04}. 
Nuclei are almost transparent for low-energy  pions, which means that invariant-mass 
distributions of pion pairs produced close to the production threshold are much less 
affected by FSI than at higher energies. Photon induced production of pion pairs off nuclei 
has been previously measured with the TAPS detector at MAMI \cite{Messchendorp_02,Bloch_07}. 
These experiments found for heavy nuclei
a systematic shift of the invariant mass of $\pi^0\pi^0$ pairs towards small values 
while the distributions for mixed-charge pairs $\pi^0\pi^{\pm}$ were much less effected. 
However, as discussed in \cite{Bloch_07} the spectra were reproduced by the results 
from calculations with the Boltzmann-Uehling-Uhlenbeck (BUU) transport model \cite{Buss_06}.
This model treated the pion-nucleus FSI effects in great detail, but did not include 
any explicit in-medium modification of $\pi^0\pi^0$ pairs. However, the model 
results had also non-negligible systematic uncertainties, mainly from the input for the
elementary reaction cross sections.

A more systematic study was limited by the statistical quality of the data and 
the range of nuclear masses investigated. As discussed above, FSI effects are 
minimized close to the production threshold; but the production cross sections 
are small. The lightest nucleus included in the previous study was $^{12}$C. 
Invariant-mass distributions for the elementary processes off the (quasi)-free 
nucleon (see \cite{Krusche_03} for an overview)
have been studied previously for the $p\pi^0\pi^0$ \cite{Haerter_97,Wolf_00,Sarantsev_08}, 
the $n\pi^0\pi^+$ \cite{Langgaertner_01}, and $p\pi^0\pi^-$ \cite{Zabrodin_99} final states.
Data for $\gamma d\rightarrow X\pi^0\pi^0$ had low statistical quality
in the threshold region \cite{Kleber_00} and no data were available for 
$\gamma d\rightarrow X\pi^0\pi^{\pm}$.  
  
The present experiment therefore aimed at a precise measurement of pion-pion
invariant-mass distributions for the $\pi^0\pi^0$ and $\pi^0\pi^{\pm}$
final states at low incident photon energies over a large range of nuclear masses, 
including $^{2}$H, $^{7}$Li, $^{12}$C, $^{40}$Ca, and $^{\rm nat}$Pb targets.
Such data are needed to separate the FSI effects from non-trivial in-medium 
modifications of the double pion channels.  

\section{Experimental setup}

The experimental setup is described in \cite{Yasser_13} where the data 
from the $^7$Li target have been analyzed for other meson production reactions. 
The measurement was performed at the tagged photon beam of the Mainz MAMI accelerator 
\cite{Herminghaus_83,Walcher_90}. The primary electron beam of 883 MeV produced 
bremsstrahlung in a copper radiator of 10$~\mu$m thickness. The photons were tagged 
with the Glasgow magnetic spectrometer \cite{McGeorge_08}. Four solid-state targets
(thicknesses 5.4~cm (Li), 1.5~cm (C), 1~cm (Ca), 0.05 cm (Pb)) and a liquid-deuterium 
target (thickness 4.8~cm, 0.6\% of radiation length) were used. The thickness of the 
solid-state targets was chosen such that they were comparable in units of the 
radiation lengths of the materials (from 3.5\% for lithium to 9\% for lead). 
For the measurement with the lead target a 
lower beam energy was used (645 MeV) in order to enhance the statistical quality  of 
the data in the threshold region. In this Letter, we therefore summarize only the 
results for incident photon energies below 600~MeV.
The reaction products were detected with an almost $4\pi$-covering electromagnetic
calorimeter assembled from the Crystal Ball (CB) \cite{Starostin_01} and TAPS 
\cite{Gabler_94} detectors. The targets were mounted in the center of the 
Crystal Ball. A Particle Identification Detector (PID) \cite{Watts_04} 
was mounted around the targets in cylindrical geometry for identification of charged 
particles. For the same purpose the TAPS detector was equipped with individual plastic 
detectors in front of each BaF$_2$ module.

The trigger conditions were chosen relatively open in order to minimize systematic 
uncertainties. The Crystal Ball and TAPS were subdivided into logical sectors, TAPS into 
8$\times$64 modules arranged in a pizza-like geometry and the Crystal Ball into 45 rectangles. 
Events were accepted when at least two sectors of the calorimeter were hit and the total 
analog sum energy in the Crystal Ball was above 50 MeV. The charged pion was not used 
in the trigger. Events where the decay photons of the neutral pions had not satisfied 
the trigger condition were removed in the off-line analysis because of the systematic 
uncertainties in the simulation of the trigger efficiency for charged pions. 
 
\section{Data analysis}

The analysis of data measured with the above setup is discussed for different final 
states in \cite{Yasser_13,Schumann_10,Zehr_12}. The specific analysis steps for 
$\pi^0\pi^0$ and $\pi^0\pi^{\pm}$ production off nuclear targets are described 
in detail for the previous measurements with the TAPS detector \cite{Bloch_07}. 
The main improvements of the present experiment compared to the previous measurements
\cite{Messchendorp_02,Bloch_07} are the almost $4\pi$ coverage of the solid angle and 
the use of the PID for the identification of charged pions via the $\Delta E-E$ method.

The first step of the analysis was the assignment of hits to photons, charged pions, 
protons, and neutrons. The identification of hits in TAPS (see \cite{Bloch_07} for details)
used the information from the plastic scintillators in front of the TAPS crystals and  
a pulse-shape analysis (PSA), using the narrow- and wide-gate energy integration of the 
BaF$_2$ signals for the separation of photons from nucleons. A time-of-flight (ToF) 
versus energy analysis showed the characteristic bands for photons, charged pions, 
and nucleons. As discussed in \cite{Bloch_07}, at forward angles the charged-pion band 
in TAPS is contaminated with protons. Therefore, charged pions were not accepted in 
TAPS (this excluded charged pions with polar angles $\Theta_{\pi}< 20^{\circ}$ from 
the analysis). In the Crystal Ball \cite{Schumann_10,Zehr_12}, charged hadrons were 
\begin{figure}[thb]
\resizebox{0.23\textwidth}{!}{%
  \includegraphics{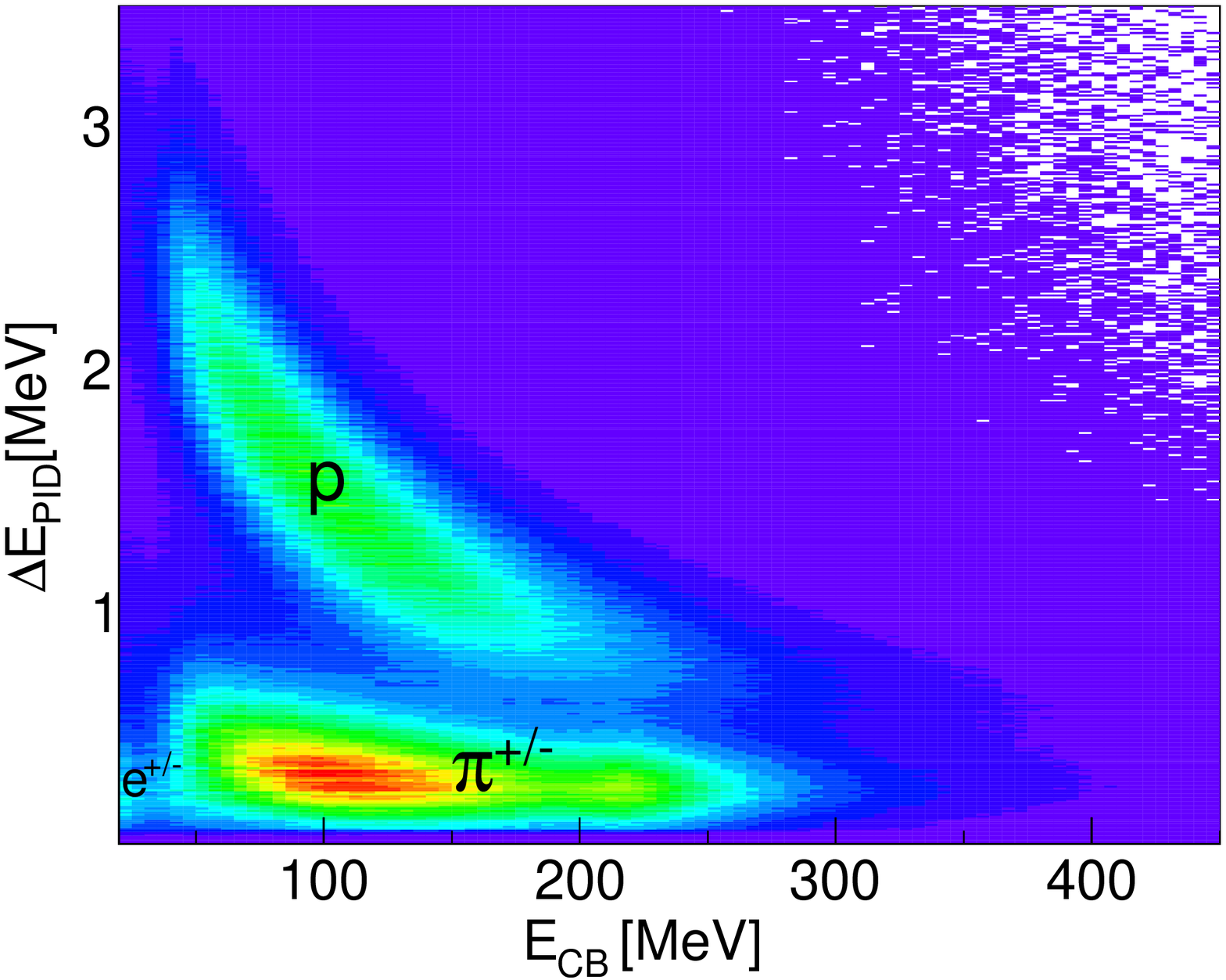}
}
\resizebox{0.26\textwidth}{!}{%
  \includegraphics{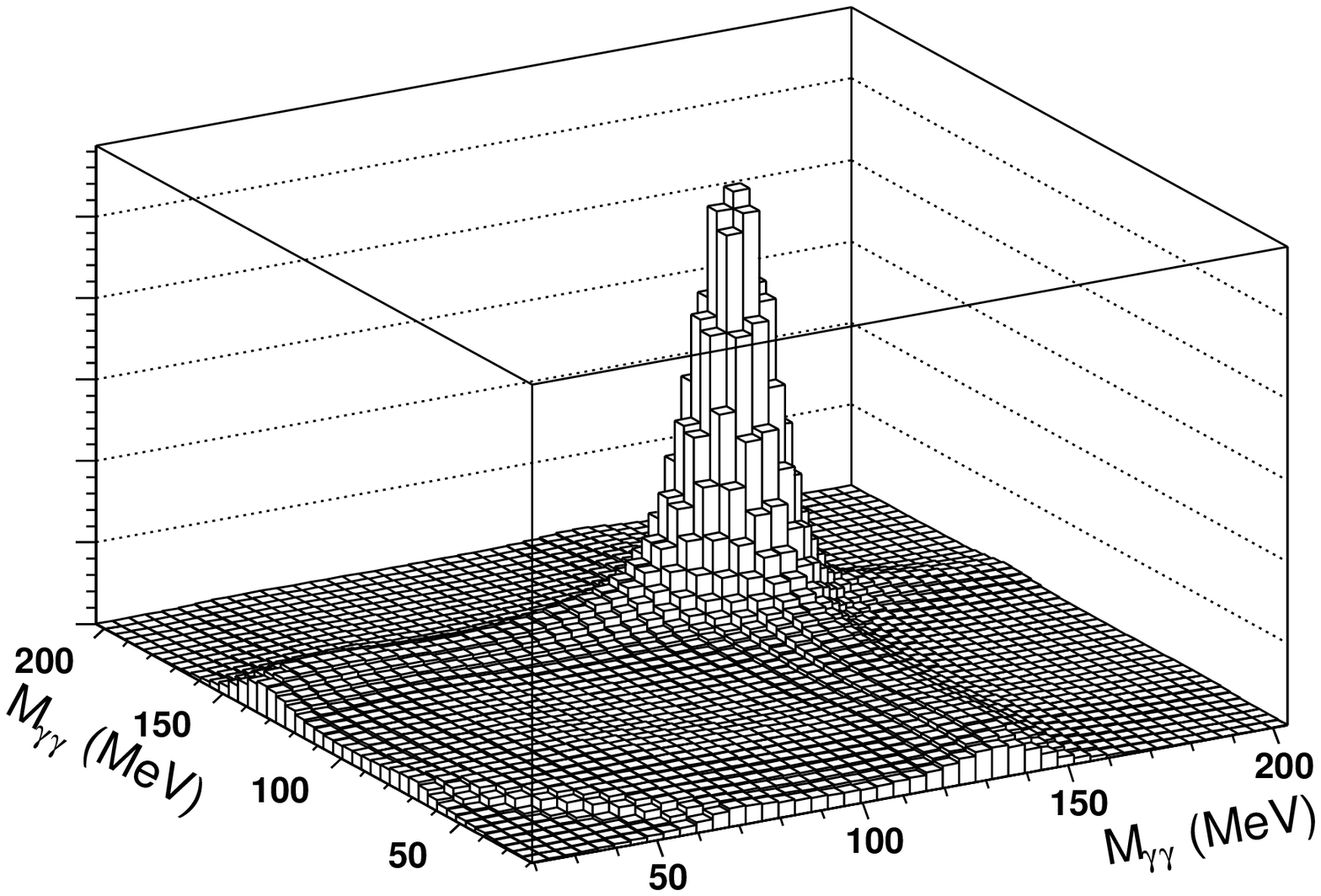}  
}
\caption{Left hand side: Energy deposition in CB versus PID. Charged pions are separated from 
protons and low energy electrons (data from Li target).
Right hand side: Two-dimensional distribution (data from Li-target) of the invariant 
masses from the $\pi^0\pi^0$ channel for incident photon energies between 
400 - 460 MeV (only `best' combinations, see text).
}
\label{fig:banana}       
\end{figure}
identified via their $\Delta E-E$ spectra constructed from the energy loss $\Delta E$ 
in the PID and the energy deposition in the Crystal Ball. This analysis also used the 
correlation between the azimuthal angles of the hits in the PID and the CB. 
A typical $\Delta E-E$ spectrum shown in Fig.~\ref{fig:banana}, left hand side,
demonstrates the separation of charged pions from protons.
In the CB, neutrons cannot be distinguished from photons, as opposed to in TAPS. 
Therefore, neutral hits in the CB were accepted as candidates for photons and for 
neutrons, while neutral hits in TAPS were separated at this stage into disjunct 
samples of photons and neutrons through PSA and ToF-versus-energy.  
 
Events were then assigned to the $\pi^0\pi^0$ and $\pi^0\pi^{\pm}$ final states. 
Accepted were events with four photons (for the $2\pi^0$ final state) and
events with two photons and one charged pion (for the $\pi^0\pi^{\pm}$ final state).
Detection of recoil nucleons was allowed but not required and after identification,
recoil nucleons were ignored in the further analysis.  
Consequently, three different sub-classes of events were accepted for both reactions.
For the double neutral channel, these were: events with exactly four photon candidates
(50\%); events with four photons and one proton (32\%); and events with four photons 
and one neutron candidate (18\%). Similarly, for $\pi^0\pi^{\pm}$ pairs, events with: 
exactly two photon candidates and exactly one candidate for a charged pion in the 
Crystal Ball (59\%); events with an additional proton (28\%); and events with an additional 
candidate for a neutron (22\%) were accepted. 

\begin{figure}[thb]
\centerline{
\resizebox{0.45\textwidth}{!}{%
  \includegraphics{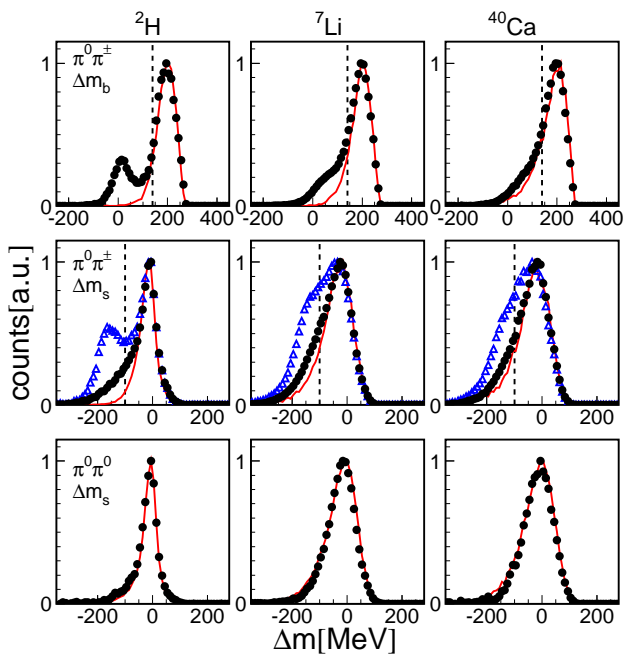}
}}
\caption{Missing mass spectra for $E_{\gamma}$ between 400 MeV and 500 MeV for $^2$H, $^7$Li,
and $^{40}$Ca. Top row: missing mass for $\pi^0\pi^{\pm}$ candidates 
for the background assumption of the $\pi^0 p$ final state. Background from single $\pi^0$
production with protons misidentified as charged pions peaks around zero. Dashed lines 
indicate cut for further analysis. Middle row: missing mass for $\pi^0\pi^{\pm}$ candidates 
for the assumption of the $\pi^0\pi^{\pm}$ final state. Signal around zero. (Blue) triangles: 
raw spectra, black dots: after cut on $\Delta m_b$. Bottom row: missing mass for the 
$\pi^0\pi^0$ hypothesis for double neutral meson pairs. (Red) solid lines: simulated 
line shape of signal.} 
\label{fig:mismas}       
\end{figure}

In the next step the invariant mass of the pairs of photon candidates was constructed. 
For events with a neutron candidate in the Crystal Ball (which is indistinguishable from 
a photon), all possible pion combinations of neutral hits were tested and the `best' 
combination was selected using a $\chi^2$-test minimizing
\begin{equation}\label{chi2-method1}
\chi^{2} = \sum_{i=1}^{n_{\gamma\gamma}}\left 
(\frac{m_{\pi^0}-m_{\gamma\gamma,i}}{\Delta m_{\gamma\gamma,i}}\right)^{2} ,
\end{equation} 
where $m_{\pi^0}$ is the nominal pion mass, and $m_{\gamma\gamma,i}$,
$\Delta m_{\gamma\gamma,i}$ are invariant mass and uncertainty of all possible
combinations of `photon' pairs. The unpaired neutral hit, not assigned to be a pion 
decay photon, was then taken as neutron and thereafter disregarded.
The resulting two-dimensional spectrum of pion-invariant masses for double $\pi^0$ 
production is shown in Fig.~\ref{fig:banana}, right hand side.
Events with both invariant masses in the range 110 - 160 MeV were selected for
further analysis. The small combinatorial background underneath the peak, was 
extrapolated in a side-bin analysis and subtracted. Subsequently, the nominal mass 
of the pion  $m_{\pi^0}$  was used to improve the resolution as in \cite{Bloch_07,Zehr_12} 
by replacing the measured photon energies $E_{1,2}$ by
\begin{equation}\label{eq:Constrain}
E_{1,2}^{'} = E_{1,2}\frac{m_{\pi^0}}{m_{\gamma\gamma}}\; ,
\end{equation}
where $m_{\gamma\gamma}$ is the invariant mass corresponding to the measured photon
four vectors. This data sample was tested for residual background with missing mass 
spectra (the recoil nucleon was treated as a missing particle even if it was detected):
\begin{equation}
\Delta m_s = \left|{\mbox{ P}_{\gamma}+\mbox{ P}_{N}
             -\mbox{ P}_{\pi_1}-\mbox{ P}_{\pi_2}}\right|
-m_N\ ,
\label{eq:misma_1}
\end{equation}
where $\mbox{ P}_{\gamma}$ and $\mbox{ P}_{N}$ are the four momenta of the incident
photon and nucleon (the latter assumed at rest) and $\mbox{ P}_{\pi_i}$, $i=1,2$ the four 
momenta of the two pions. The corresponding spectra are shown in Fig.~\ref{fig:mismas}, bottom
row. They demonstrate the cleanness of the signal and the excellent agreement between measured
data and Monte Carlo simulations.

For the mixed-charge channel, the invariant-mass spectra were fitted with the 
simulated line-shape of the pion peak and a polynomial background (second degree polynomial). 
Some background arises from the quasi-free $\gamma p\rightarrow p\pi^0$ reaction from 
protons which contaminate the pion band in the $\Delta E-E$ spectra. The fraction of protons 
that intrude is small, but the cross section for single $\pi^0$ production can be larger 
than for $\pi^0\pi^{\pm}$ production by more than two orders of magnitude. The most efficient 
way to remove this background uses the missing mass calculated for the hypothesis of the 
$\gamma p\rightarrow p\pi^0$ reaction (see \cite{Bloch_07,Zehr_12} for details), i.e. it 
is assumed that the `charged pion' is a misidentified proton:
\begin{equation}
\Delta m_b = \left|{\mbox{ P}_{\gamma}+\mbox{ P}_{N}
             -\mbox{ P}_{\pi^0}}\right|
-m_N\ ,
\label{eq:misma_2}
\end{equation}
with the same notation as in Eq.~(\ref{eq:misma_1}). Typical spectra are shown in 
Fig.~\ref{fig:mismas}. For the deuteron target background peak and double pion production are
separated, for the heavier nuclei, due to the larger momenta of the bound nucleons and 
FSI, the background is only visible as a shoulder on the tail of the signal peak. 
Only events with $\Delta m_b >$ 140~MeV have been accepted. The center row of 
Fig.~\ref{fig:mismas} shows the missing mass $\Delta m_s$ (see Eq.~(\ref{eq:misma_1})) 
calculated under the assumption of quasi-free $\pi^0\pi^{\pm}$ production. The (blue) 
triangles represent the raw spectra and the (black) dots the events that passed the 
above $\Delta m_b$ cut. The latter agree well with the simulated line-shape ((red) lines). 
Residual background was removed with the condition $\Delta m_s > -100$~MeV.   

Charged pions may deposit more than their kinetic energy in the detector. 
When they are stopped, they may interact with nuclei or decay and deposit some 
fraction of their rest mass as energy (see \cite{Bloch_07} for details). 
The additional energy deposition was taken into account by a correction derived from 
the comparison of kinetic and deposited energies of charged pions in the Monte Carlo 
simulations. The charged pion reconstruction was tested as in \cite{Bloch_07} with an 
analysis of the position and shape of the invariant mass peak from 
the $\eta\rightarrow \pi^0\pi^+\pi^-$ decay. The results for data and Monte Carlo 
simulation were in good agreement, and the peak position agreed with the nominal 
mass of the $\eta$-meson. Since charged pion detection enters twice in this analysis, 
this test is very sensitive. 

Total cross sections were extracted from the measured yields, the target surface
densities, the photon flux, and the simulated detection efficiencies. 
Systematic uncertainties from the targets were in the range from 1\% to 3\%.
The photon flux was determined from the counting of the deflected electrons in the 
tagger focal plane and the tagging efficiency, i.e. the fraction of correlated 
photons passing the collimator (typically $\approx$ 50\%). The systematic
uncertainty was estimated at the 5\% level. Systematic uncertainties from the cuts
applied for reaction identification and possible residual backgrounds
were estimated at $\approx$ 3\% for $\pi^0\pi^0$ and $\approx$ 5\% (for $E_{\gamma}=$ 500 MeV) 
to 10\% (for $E_{\gamma}=$ 400 MeV) for $\pi^0\pi^{\pm}$.
The detection efficiency was simulated with the Geant4 program package \cite{GEANT4}.
The simulations were based on the event generator used for the previous experiments 
\cite{Messchendorp_02,Bloch_07}, which incorporates Fermi-smearing and FSI processes in a 
Glauber-type approximation. 

The efficiency corrections were applied to the data as a function of incident photon 
energy and invariant mass of the pion pairs. In this way, the influence of the properties 
of the event generator on the results is minimized. Depending on incident photon energy, 
pion-pion invariant mass, and target, typical values of the detection efficiency 
were in the range 30\% - 70\% for $\pi^0\pi^0$ pairs and between 5\% - 40\% for the 
$\pi^0\pi^{\pm}$ channel. Total cross sections were extracted by integration of the 
invariant mass distributions. We estimate a systematic uncertainty of the detection 
efficiency of $\approx$~10\% for neutral pairs and $\approx$ 20\% for the mixed 
charge pairs. It is larger for the latter because their detection efficiency depends
strongly on the kinetic energy of the charged pions (see \cite{Zehr_12} for details). 
The total systematic uncertainties are indicated in Fig.~\ref{fig:invdiff} 
as shaded bands (error bars in all figures represent statistical uncertainties). 
One should, however, keep in mind that for the comparisons of 
reaction channels and the results for different nuclei, systematic uncertainties 
cancel to a large extent.

\section{Results}

\begin{figure}[thb]
\centerline{
\resizebox{0.45\textwidth}{!}{%
  \includegraphics{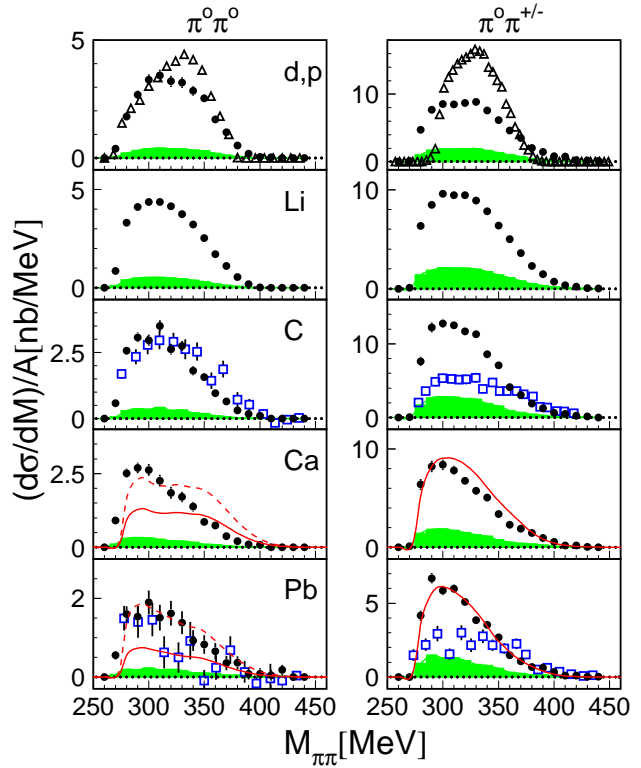}
}}
\caption{Invariant mass distributions of pion pairs normalized to the nuclear 
mass numbers for incident photon energies 400 - 460 MeV. 
From top to bottom for deuterium (and proton \cite{Zehr_12}, open triangles), 
lithium, carbon, calcium, and lead targets.
Open (blue) squares: previous data from \cite{Messchendorp_02}.
Shaded (green) bands: systematic uncertainties. (Red) solid curves: GiBUU results
\cite{Buss_12}, dashed: arbitrary rescaled for easier comparison of the shapes.} 
\label{fig:invdiff}       
\end{figure}

Invariant-mass distributions of the pion pairs for incident photon energies
between 400 MeV - 460 MeV are summarized in Fig.~\ref{fig:invdiff}.
The distributions measured for pion pairs produced off quasi-free nucleons from the 
deuteron differ in shape and magnitude from those measured off free protons 
(Fig.~\ref{fig:invdiff}, top row). This may be partly attributed to
Fermi-smearing, but includes also differences in the elementary cross sections 
for protons and neutrons. This is important for comparisons of the nuclear data 
to model results, which have to rely on experimental input for the elementary 
cross sections.   

\begin{figure}[thb]
\centerline{\resizebox{0.4\textwidth}{!}{%
  \includegraphics{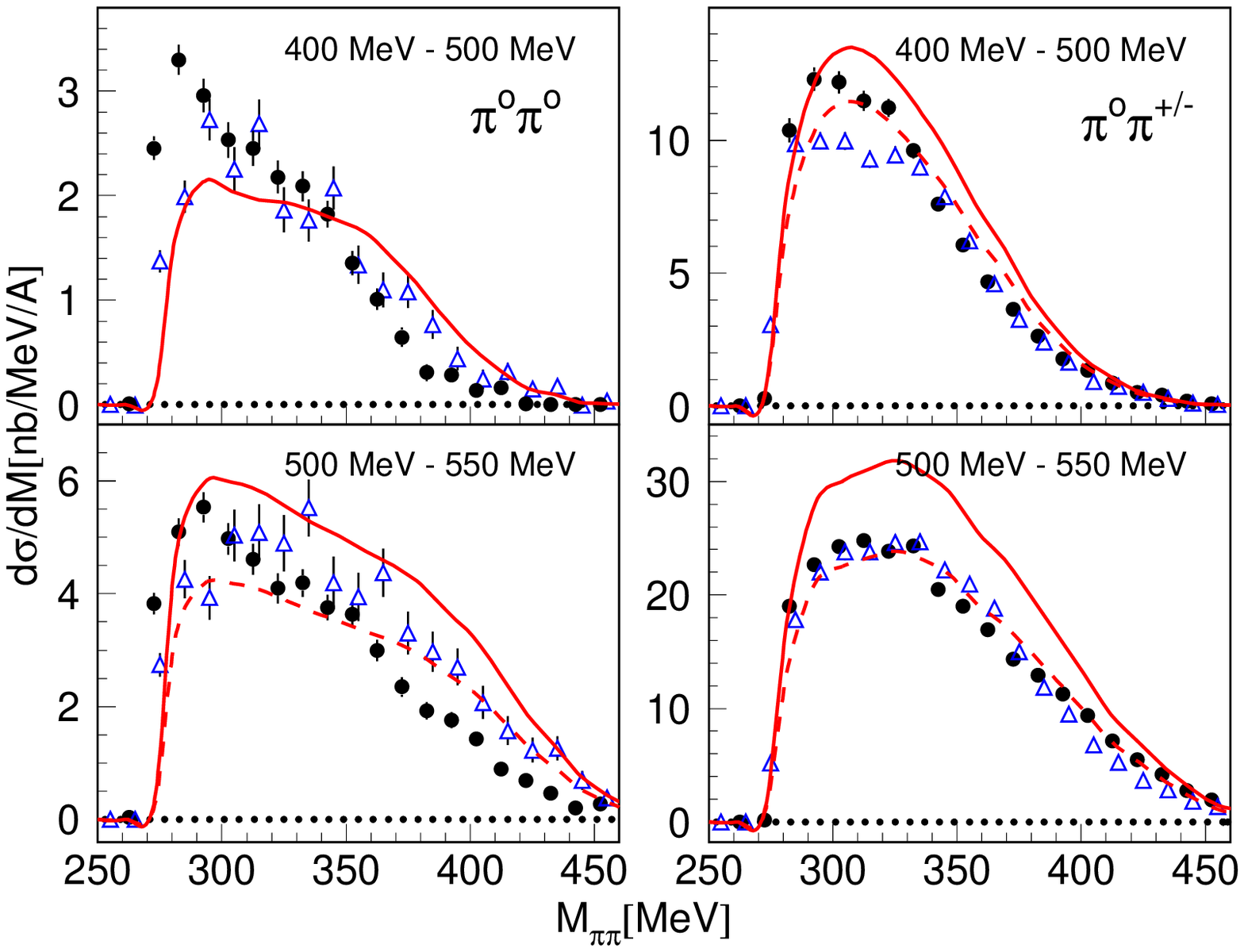}
}}
\centerline{\resizebox{0.4\textwidth}{!}{%
  \includegraphics{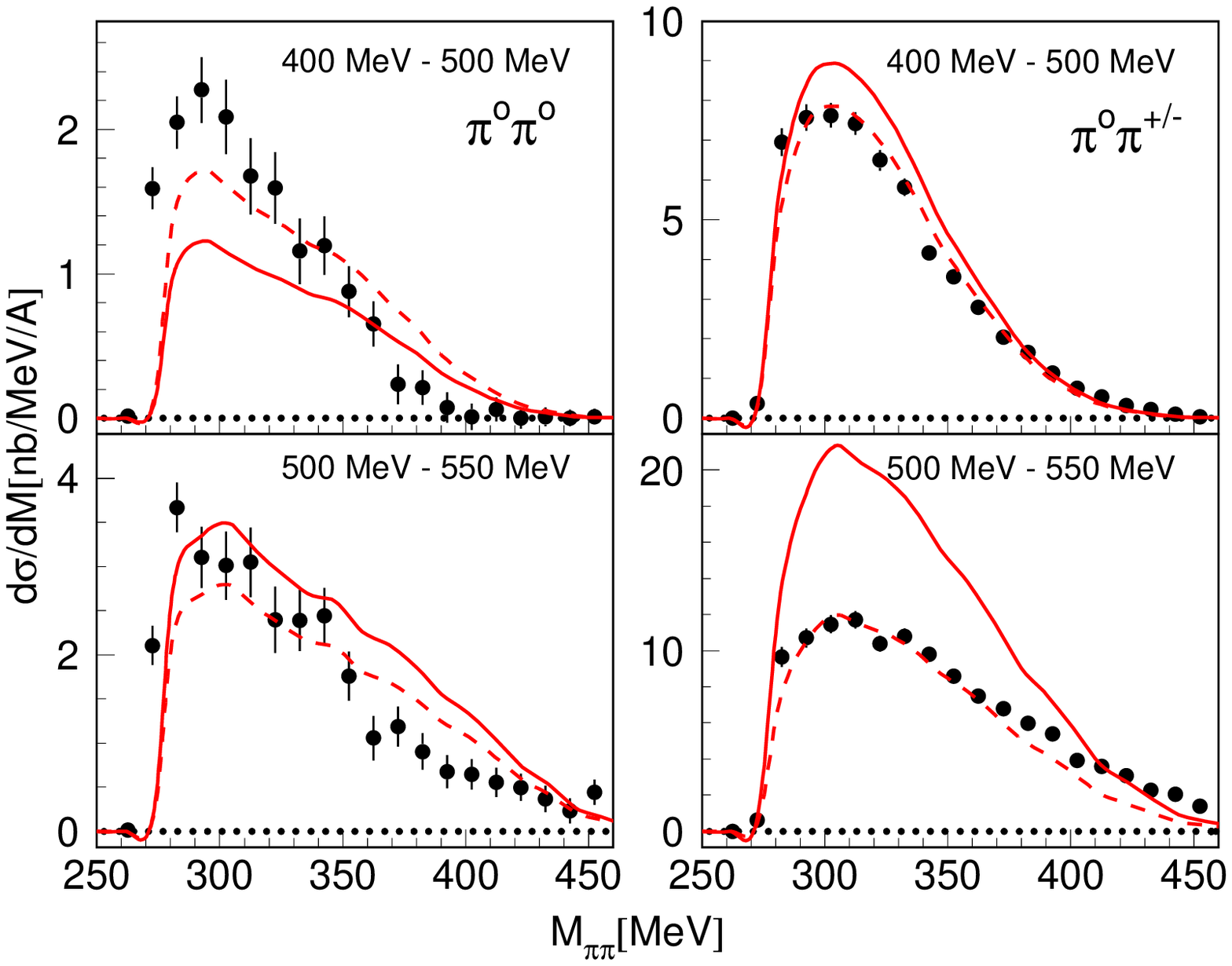}
}}
\caption{Top: Invariant-mass distributions of pion pairs for $^{40}$Ca normalized to 
the nuclear mass numbers. Left hand side: $\pi^0\pi^0$, right hand side:
$\pi^0\pi^{\pm}$. upper row: incident photon energies from 400 MeV - 500 MeV,
lower row: 500 MeV - 550 MeV (black dots). (Blue) triangles: previous data from 
Bloch et al. \cite{Bloch_07}, (red) solid curves: GiBUU model results from Buss et al. 
\cite{Buss_06} (dashed curves: same results arbitrary rescaled to data for better
comparison of shape). Bottom: same distributions from present data for lead.} 
\label{fig:invdiff2}       
\end{figure}

The data for carbon and lead from Messchendorp et al. \cite{Messchendorp_02}
are in reasonable agreement with the present results for the neutral pairs; 
but not for the mixed-charge channel. Agreement between the $\pi^0\pi^{\pm}$ data 
for calcium from Bloch et al. \cite{Bloch_07} and the present results is much better
(see Fig.~\ref{fig:invdiff2}). It had already been noticed that the 
$\pi^0\pi^{\pm}$ data from references \cite{Messchendorp_02,Bloch_07} are 
probably in conflict because the mass dependence for the chain 
carbon - calcium - lead would have been very unnatural (the magnitude of the calcium
cross section normalized by mass number had been larger than both the carbon and 
lead data, but one would have expected it in between of them). The present 
measurement favors the previous results from Bloch et al. \cite{Bloch_07} with
which they agree within systematic uncertainties. 

\begin{figure}[thb]
\centerline{
\resizebox{0.45\textwidth}{!}{%
  \includegraphics{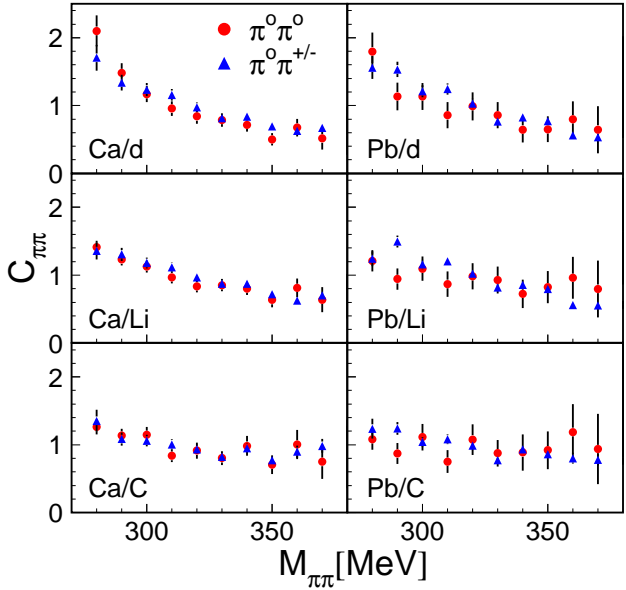}
}}
\caption{Composite ratios for $\pi^0\pi^0$ (red dots) and  $\pi^0\pi^{\pm}$ (blue
triangles) for incident photon energies between 400 MeV - 460 MeV. Left hand side:
calcium, right hand side: lead. From top to bottom compared to d, Li, C.}
\label{fig:ratio}       
\end{figure}

Comparison of the invariant-mass distributions in Fig.~\ref{fig:invdiff} shows, that
for the $\pi^0\pi^0$ and the $\pi^0\pi^{\pm}$ pairs, strength shifts to small
invariant masses for increasing mass of the nucleus. This general trend is certainly
not related to properties of the $\sigma$ meson, which does not couple to the 
mixed-charge channel. The effect can be studied in more detail with the help of the
composite ratios defined by  
\begin{equation}
C_{\pi\pi}(A_{1},A_{2}) = \frac{(d\sigma(A_{1})/dM) 
/ \sigma(A_{1})}{(d\sigma(A_{2}) / dM) / \sigma(A_{2})}
\end{equation}
where $d\sigma(A_{i})/dM$ are the invariant mass distributions and $\sigma(A_i)$ the 
total production cross sections for two nuclei with mass numbers $A_{1}$ and $A_{2}$.    
Results for incident-photon energies from 400 MeV - 460 MeV are summarized in 
Fig.~\ref{fig:ratio}. They demonstrate that, at least in this energy range, the behavior 
of the invariant-mass distributions is almost identical for both isospin states, so that 
the dominant effect is most likely due to FSI of the pions with the nucleus. 
Previous results \cite{Messchendorp_02} for the composite ratio of carbon and lead 
indicated a stronger enhancement at low invariant masses for neutral pion pairs than 
for the mixed-charge pairs. However, comparison to the present 
invariant-mass distributions (see Fig.~\ref{fig:invdiff}) suggests that this was 
due to statistical fluctuations, in particular in the lead data. Using the lighter nuclei
(deuteron and $^7$Li) as reference, the low-mass enhancement becomes larger, but in
the same way for neutral and mixed-charge pairs.

\begin{figure}[thb]
\centerline{
\resizebox{0.45\textwidth}{!}{%
  \includegraphics{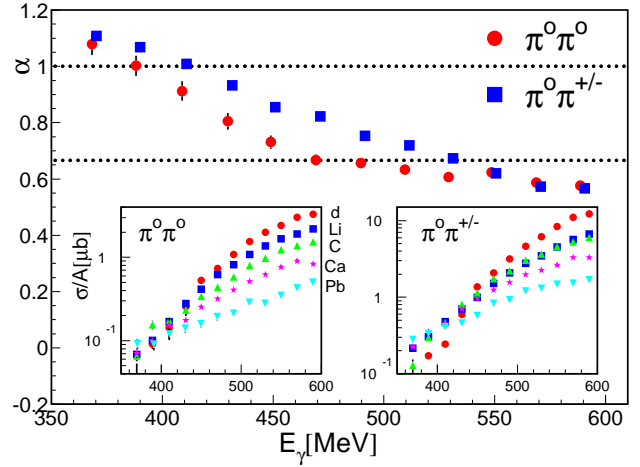}
}}
\caption{Scaling behavior of total cross sections. 
Main figure: Scaling coefficient $\alpha$ as a function of incident photon energy.
(Red) dots: neutral pairs, (blue) squares: mixed-charge pairs (symbols slightly
displaced for better readability) 
Inserts: total cross sections scaled by $A$ as a function of incident photon energy.
}
\label{fig:alpha}       
\end{figure}

An estimate of how important FSI effects are for the pions can be deduced from the scaling 
of the cross sections with nuclear mass number $A$. Total cross sections as a function 
of incident photon energy, scaled by the mass number $A$, are shown for both isospin states 
in the inserts of Fig.~\ref{fig:alpha}. The scaling changes from threshold to higher 
incident photon energies. This behavior can be parameterized with the scaling coefficient 
$\alpha$ from
\begin{equation}
\label{eq:alpha}
\sigma (A,E_{\gamma})\propto A^{\alpha (E_{\gamma})}\;,
\end{equation}
which is shown in Fig.~\ref{fig:alpha}. Their interpretation is straightforward. 
At incident photon energies close to the production threshold $\alpha$ is close to 
unity, which means that the cross sections scale with the number of nucleons 
(or the nuclear volume), indicating negligible losses due to pion absorption. 
The coefficient then drops as a function of $E_{\gamma}$ and approaches 2/3 for beam 
energies between 500 MeV and 600 MeV, indicating a scaling proportional to the 
surface of the nuclei, which means strong absorption. This is expected from the 
absorption properties of nuclear matter for pions \cite{Krusche_04}. Nuclei are 
transparent for pions with kinetic energies below $\approx$ 40 MeV and `black' for 
pions with energies above $\approx$~100 MeV, which may excite the $\Delta$-resonance. 
In quasi-free kinematics, pions from pion pairs produced with incident photon energies
around 500 MeV (600 MeV) have kinetic energies around 50 MeV (75 MeV) when the energy 
is symmetrically shared by the two pions (up to 138 (150) MeV for pions from 
extremely asymmetric energy distribution). The scaling behavior is similar for both 
isospin channels, so that comparable FSI effects are to be expected. 

\begin{figure}[thb]
\centerline{\resizebox{0.45\textwidth}{!}{%
  \includegraphics{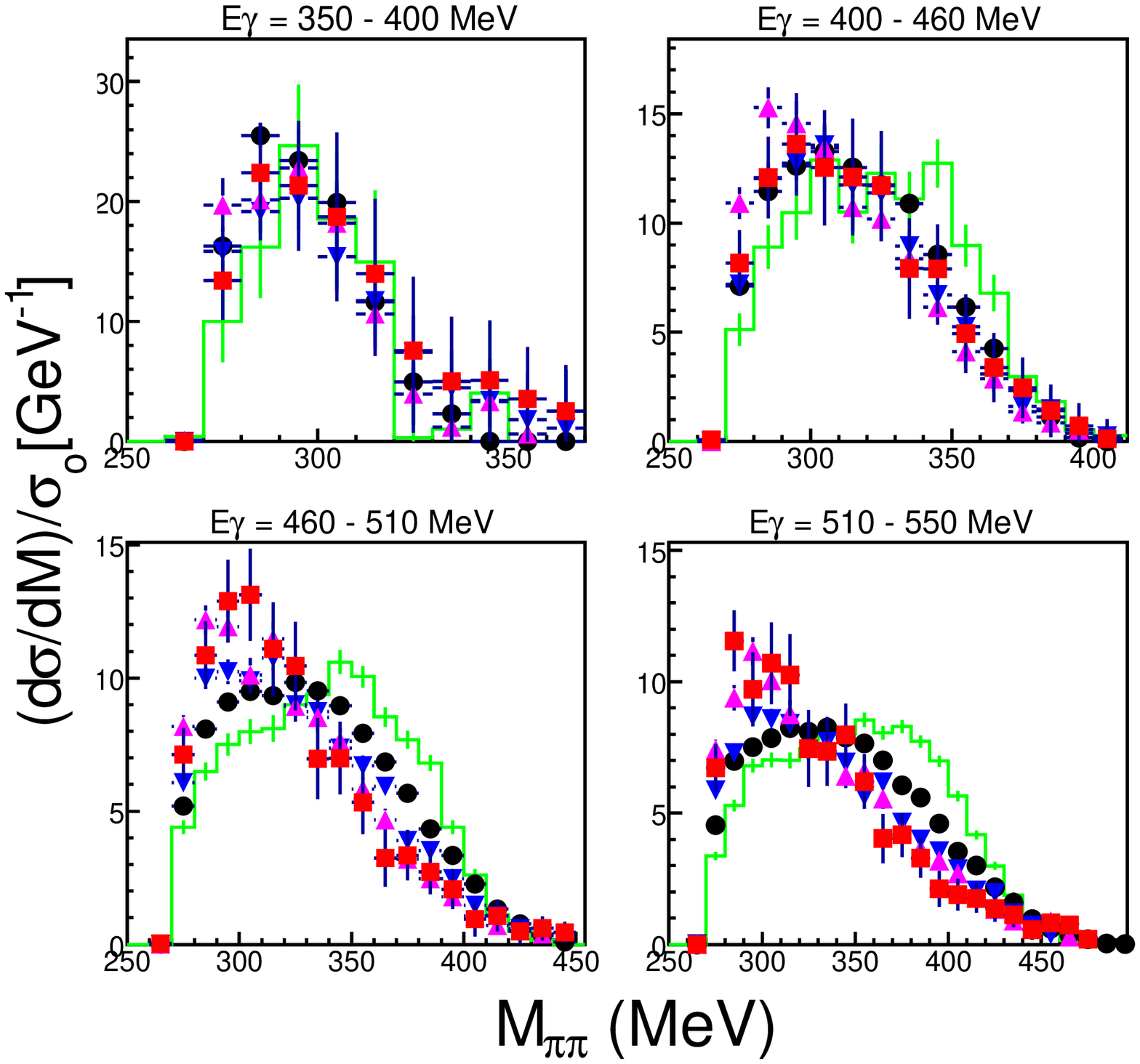}}}
\centerline{\resizebox{0.45\textwidth}{!}{%
  \includegraphics{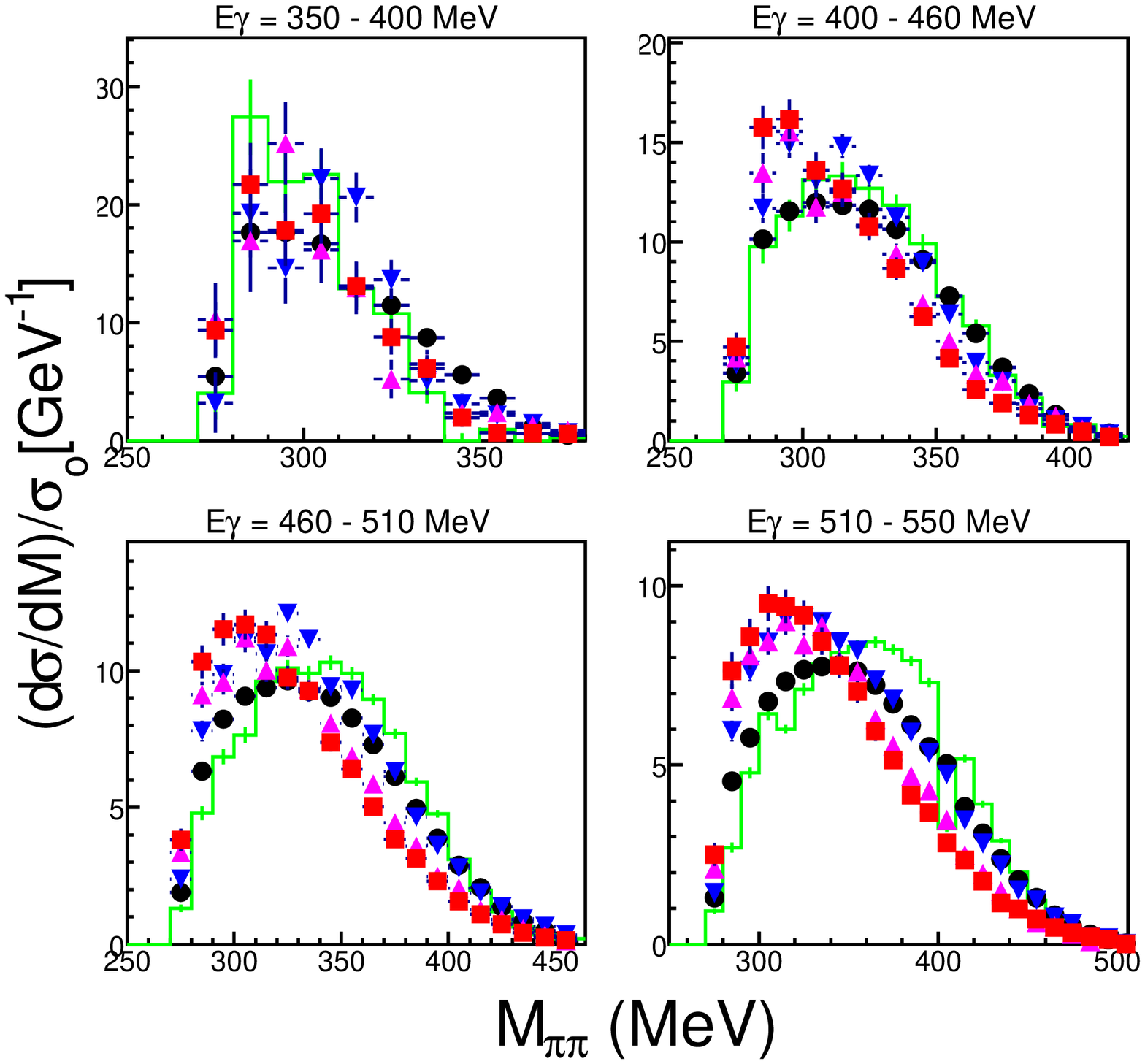}
}}
\caption{Pion-Pion invariant mass distributions for different ranges of incident 
photon energies normalized to the total cross sections.
Top part: $\pi^0\pi^0$, bottom part: $\pi^0\pi^{\pm}$.
Deuteron: (green) histogram, Li (black) dots, C (blue) down triangles, 
Ca (magenta) up triangles, Pb (red) squares).}
\label{fig:compinv}       
\end{figure}

The shapes of pion-pion invariant-mass distributions for different ranges of incident 
photon energies are compared in Fig.~\ref{fig:compinv}. For both isospin channels,
at the lowest incident photon energies, the distributions are similar for all nuclei.
They basically agree with a Fermi-smeared version of the elementary cross section
average of proton and neutron cross sections, represented by the deuteron data.
At higher incident photon energies they start to differ, and for fixed photon energy, 
the heaviest nuclei have the softest distributions. This general trend can be related 
to the increase of FSI with rising beam energy and rising mass. It can be easily 
understood that FSI tends to soften the distributions. Two effects are important. 
Inelastic scattering tends to decrease the kinetic energy of the pions and their 
mean-free path increases with decreasing energy. 

Any effects due to in-medium modification of the $\sigma$-meson can only be studied
on top of the FSI effects, which requires detailed model descriptions of FSI.
An efficient way to treat these `trivial' nuclear effects are transport-theoretical
approaches \cite{Buss_12}. We compare typical invariant-mass distributions to the
results of the Giessen Boltzmann-Uehling-Uhlenbeck (GiBUU) model \cite{Buss_12} in
Figs.~\ref{fig:invdiff} and \ref{fig:invdiff2}. On an absolute scale, the production cross
section for the neutral pairs is underestimated by the model at very low incident
photon energies and agrees better at higher energies, while for the mixed-charge pairs
agreement is good at low energies and becomes worse for the highest energies.
These discrepancies are at least partly related to the uncertainty in the elementary
input to the model (the production cross sections off neutrons are not well known
for most isospin channels). In order to facilitate the comparison of the shapes,
rescaled versions of the model results are shown. The shapes of the invariant-mass
distributions of the mixed-charge pairs agree for all discussed ranges of incident
photon energy almost perfectly with the data. This demonstrates that for $\pi^0\pi^{\pm}$
the observed evolution of the distributions with photon energy and atomic mass can 
be indeed explained by FSI. Agreement is not as good for the neutral pairs where, at
least for low incident photon energies, the measured distributions tend to be softer
than the predicted ones, exhibiting additional strength at low invariant masses. 
However, this effect is not large and part of it is probably also related to the
model input. The shape of the distribution measured for the deuteron 
target is softer than for the proton (see Fig.~\ref{fig:invdiff}) and this effect is 
not included in the model.       

\section{Summary and Conclusions}	

Precise results have been obtained for the invariant-mass distributions of
$\pi^0\pi^0$ and $\pi^0\pi^{\pm}$ pairs produced from deuterium, Li, C, Ca, and Pb
targets from production threshold up to incident photon energies of 600 MeV.
A pronounced shift of strength towards small invariant masses as a function of the 
nuclear mass number is observed for {\em both} final states. This effect increases 
with increasing beam energy and can be related to re-absorption and re-emission 
processes of the pions in the nuclei. An analysis of the scaling behavior of the 
total cross sections demonstrates that FSI are negligible close to production threshold 
and saturate at energies above 500 MeV. The general trend of the
shape change of the invariant-mass distributions correlates with the energy 
dependence of the FSI effects. Consequently, the dominant effect on the invariant-mass
distributions comes from FSI. As a consequence of these results,
possible in-medium modifications of the $\sigma$ meson (respectively pion-pion
interaction in the scalar, isoscalar state) cannot be easily tested with an analysis 
of the shape change of the pion-pion invariant-mass distributions as a function 
of mass number and/or beam energy as attempted in many recent works.  
Predictions of the invariant-mass distributions in the framework of the GiBUU 
transport model reproduce their shapes for the mixed-charge pairs excellently,
demonstrating that the nuclear effects for this channel are understood.
Small shape discrepancies for the neutral pairs require further investigation, in
particular more precise input for the elementary reactions into models.
In view of these uncertainties, no clear evidence for a contribution of 
$\sigma$ in-medium modification can be deduced, although the observed shapes of
the low-energy invariant-mass distributions of $\pi^0\pi^0$ pairs suggest that a
small effect might exists.   

\section*{Acknowledgments}
We wish to acknowledge the outstanding support of the accelerator group 
and operators of MAMI. We thank O. Buss and U. Mosel for interesting
discussions and the provision of the GiBUU model predictions.
This work was supported by Schweizerischer Nationalfonds, 
Deutsche Forschungsgemeinschaft (SFB 443), 
UK Science and Technology Facilities Council, STFC,
European Community-Research Infrastructure Activity (FP6), 
the US DOE, US NSF, and NSERC (Canada).

\end{document}